\documentclass[trackchanges]{aastex631}

\newcommand{\skipthis}[1]{}

\usepackage{graphicx}
\usepackage{hyperref}
\begin{document}

\title{The Variable Radio Emission of V830 Tau 
and Its Putative Planet}

\author[0000-0001-5643-8421]{Rachel A. Osten}
\affiliation{Space Telescope Science Institute \\
3700 San Martin Drive\\
Baltimore, MD 21218 USA}
\affiliation{Department of Physics and Astronomy Center for Astrophysical Sciences\\
The Johns Hopkins University \\
3400 N. Charles Street \\
Baltimore, MD 21218 USA}

\author[0000-0002-0826-9261]{Scott J. Wolk}
\affiliation{Center for Astrophysics | Harvard \& Smithsonian \\
60 Garden Street \\
Cambridge, MA 02138 USA}

\begin{abstract}
We report on the first moderate-length time-scale observations of a young stellar object (YSO) at microwave frequencies.
V830 Tau was monitored over the course of eight days with the JVLA at a frequency range of 4-8 GHz. 
Previous brief radio observations of this purported planet-hosting star indicated a radio-bright source with sparse evidence of dramatic intensity changes.
Our observations confirm variability larger than a factor of five over the 8 days, with closer-spaced data indicating a long-lived flare event spanning multiple days.
We discuss a hypothesis that the large, long-duration radio flare may be produced as a result of magnetospheric interaction between the star and its purported planet, using multi-year monitoring
of a few active binary systems with the Green Bank Interferometer to augment our discussion.
Although we cannot disentangle the effect of the large stellar surface area from any effects of orbital separation, the
disputed star-planet system would have a large amount of power generated from stretching and breaking of magnetic fields. 
If this long-duration flare behavior repeats with additional data on timescales close to the planetary orbital period, 
microwave signatures of interacting magnetospheres
could be a new observational tool to confirm the existence of planets around young, magnetically active stars. 
\end{abstract}


\section{Introduction} \label{sec:intro}
Young stars in the cool half of the HR diagram display enhanced levels of magnetic activity \citep{FeigelsonMontmerle1999}; this is primarily due to their youth and the consequent fast rotation experienced prior to significant angular momentum loss from a magnetized wind. 
Their magnetic activity levels have generally made it more difficult to
disentangle planetary-induced radial velocity signatures from stellar activity variability \citep{Boisse2011}, although attempts in the last several years have  mitigated the activity signals with
detections \citep[e.g.][]{Donati2016,SuarezMascareno2022}. 

Because of the enhanced activity, young stars can also be radio-bright and variable \citep[e.g.][]{so,bower2003,Dzib2015,Forbrich2017}.
For stars without an accretion disk, the
radio emission is presumed to originate primarily from gyrosynchrotron emission,
due to the presence of accelerated particles gyrating around closed magnetic field lines above the surface of the star
\citep{FeigelsonMontmerle1999}.
The observational characteristics of the radio emission which lead to this conclusion are the 
observed variability and moderate levels of circular polarization \citep{Zapata2004,Rivilla2015,Forbrich2017}.
The radio observations generally tend to be sparsely sampled, with only a handful of epochs (some of which might be quite short, only a few minutes long), and a tendency of these epochs to be separated by months to years. 
The work of \citet{Forbrich2017} utilized multiple hours-long observations at microwave frequencies and was revealing; they found evidence for extreme levels of variability (greater than a factor of 10) in young stars in the Orion Nebula Complex, but even in this case,
there were only a handful of epochs. 
\citet{massi2006} used a long time-series observation of the enigmatic young star V773 Tau to determine a periodicity to the flaring (see more below).
This small number of results is in contrast with optical space-based photometry from Kepler and TESS \citep[e.g.][]{walkowicz2011,gunther2020} which provide highly sampled data (as fast as 20 s) lasting for weeks to years, and has revolutionized stellar optical photometric studies.

Recent results have revealed young stellar binaries in eccentric orbits which can have interacting magnetospheres at periastron. 
The repeated 90 GHz observations of the pre-main sequence binary system V773 Tau discussed in  \citet{massi2006}  show remarkable flaring activity around periastron passage in the slightly eccentric system. 
The flares occur periodically with recurrence timescales approximately the orbital period.
The interpretation of these phenomena put forward was a scenario in which the two stellar magnetospheres interact.
\citeauthor{massi2006} proposed that the magnetic structure was compatible with that of a helmet
streamer that, as in the solar case, can occur at the top of the X-ray-emitting, stellar-sized coronal loops of one of the stars. 
The streamer may extend up to $\sim20$ $R_{\star}$ and interact with the corona of the other star at periastron passage, causing recurring flares \citep[see Fig.~2 of ][]{Massi2008}. 
The inferred magnetic field strength at the two mirror points of the helmet streamer is in the range 0.12 - 125 G, and the derived high Lorentz factor and detection of linearly polarized emission 
\citep{phillips1996} implies synchrotron radiation rather than gyrosynchrotron. 
An additional source, DQ Tau \citep{Salter2010}, may also exhibit recurrent mm flares at periastron.


There are several lines of evidence suggesting that stars and the planets around them interact, with mechanisms related to gravitational and magnetic interactions. 
These primarily affect close-in planets, and most observations to date have focused on hot Jupiters. 
Several systems with hot Jupiter companions display modulation 
of Ca~II and/or X-rays 
at the synoptic (planetary) cadence \citep[e.g. HD~179949,  $\nu$~And, $\tau$~Boo, and HD~189733][]{Shkolnik2008,Pillitteri2015,Gurdemir2012}.  
In particular, \citet{PoppenhaegerWolk2014} suggested 
that a star-planet interaction was responsible for maintaining the rotation period of HD~189733 at a fast rate despite the relatively old age of the star.  Meanwhile, tidal interactions may limit the lifetime of WASP-12b to about 3 Myr before it is subsumed into its stellar host \citep{Patra2017}.
The timing of flares at both X-ray and optical wavelengths has been used to establish evidence of interaction between the star and its planet \citep{Pillitteri2011,Ilin2024}.
Results in the last several years have presented radio evidence of star-planet interactions, mainly through the existence of highly circularly polarized bursts.
\citet{VilladsenPineda2023} found recurring polarized bursts approximately in phase with the planetary companion to a quiescent M dwarf star YZ Cet, while \citet{Vedantham2020} used the existence of a circularly polarized burst at lower frequencies from a quiet M dwarf to suggest the possible presence of a planet in the system.  

Active binary systems are potential analogues for understanding some of the reported star-planet interactions.  
Such systems are known to display enhanced levels of magnetic activity \citep{schrijverzwaan}, including extreme flares \citep{Osten2007,Osten2016}, presumably due to the enforced fast rotation from tidal locking as well as the interacting magnetic fields of the two stars \citep{UchidaSakurai1983}.
In detached binary systems the two stellar surfaces are close to each other;  no significant mass transfer is occurring in RS CVn (generally composed of at least one evolved G/K star; ) and BY Dra systems (generally composed of main sequence stars with a K or M primary), while in semi-detached Algol systems
mass transfer is occurring (see section 2 in \citet{berdyugina2005} for more information). 
A landmark study by \citet{Richards2003} analyzed years-long radio time-series of several RS CVn, BY Dra, and Algol systems with near-daily monitoring using the Green Bank Interferometer (GBI), finding evidence of periodicities in the radio variability, which they interpreted as evidence for magnetic cycles.

V830 Tau is a diskless  0.76 M$_\odot$ pre-main sequence (PMS) K7/M0 star in the Taurus star forming region \citep{EsplinLuhman2019}. 
As such, it has an age of about 2.5 Myr and a radius of $\sim 1.8R_\odot$.  
The Gaia DR3 distance is 130.4 $\pm 0.3$ pc \citep{GaiaDR3}.
Recent K2 data show a very stable 2.75 day period 
with no other optical signatures except for possible white light flares \citep{Rebull2020}. 
The star is X-ray bright, 
with an X-ray luminosity estimated at 5.18$\times 10^{30}$ erg s$^{-1}$ and a little over $10^{31}$ erg s$^{-1}$  overall in the extreme ultraviolet band shortward of 912 \AA. 
This does not include an observed flare which was very hot ($\sim 22$ MK) and energetic \citep{Gudel2007}.
Similarly, the system is radio bright, 0.32 and 0.37 mJy at 4.5 and 7.5 GHz, respectively, with significant variability in both bands \citep{Dzib2015} based on observations spanning only $\sim$ 3 minutes per epoch. 
\citet{Bower2016} noted a detection of the system at 6 GHz of 919$\pm$26 $\mu$Jy along
with non-detections in two other epochs with upper limits of $<$66 and 150 $\mu$Jy, 
respectively. The radio exposures were very short, with only about 4 minutes per epoch. 
A redshifted CO outflow has been detected 0.28pc in length 
\citep{Li2015}, presumably a remnant from its earlier active phase (the morphology in their Figure~36 is horseshoe-shaped so not a driven flow).

\citet{Donati2016} reported that the star is orbited by a hot proto-Jupiter of about 0.77 M$_{Jup}$ with a period of 4.93 days, implying a separation from the host star of 0.057 au, and a best fit eccentricity of 0.3 (while an eccentric orbit was a better fit to the data, the authors selected a circular orbit due to the uncertainties on the derived value of eccentricity).  
This is the youngest known exoplanet.  
Later authors \citep{Damasso2020} were unable to confirm the existence of this planet,
based on different data -- \citeauthor{Donati2016}
explicitly looked at magnetically sensitive lines, combining data from two telescopes, with high signal to noise ratio (SNR) of 70-170, while \citeauthor{Damasso2020} used data from a single telescope with individual spectra having
SNR around 20, but with more measurements. 
This casts doubt on the ability to disentangle the magnetic activity signatures from planet-induced radial velocity signatures. 
Given the close-in nature of the purported planet together with its eccentric orbit, and the evidence for strong radio variability (\citeauthor{Bower2016} et al. note that the radio observations indicate that the system ``is strongly variable on a timescale of weeks with an amplitude of variability greater than 10$\times$"), the system provides an excellent  candidate to test hypotheses about star-planet interaction.
At the same time, extended radio monitoring of V830 Tau fills in a 
unique region of parameter space in understanding radio variability from young stars, apart from its putative planet.

Here we present the results of a campaign to monitor the radio flux density of V830 for a little over a week with the Jansky Very Large Array (JVLA). 
This will both expand our perspective about longer time-scale radio variability on young stellar objects as well as provide a potential probe of a new type of star-planet interaction. 
\S 2 describes the JVLA observations, \S 3 provides additional context from multi-year radio monitoring of a few active binary systems, \S 4 
presents a discussion of the findings,  \S 5 explores our hypothesis of large flares as signatures of star-planet interactions, and \S 6 concludes.

\section{JVLA observations}

\begin{figure}[htbp]
    \includegraphics[scale=0.1]{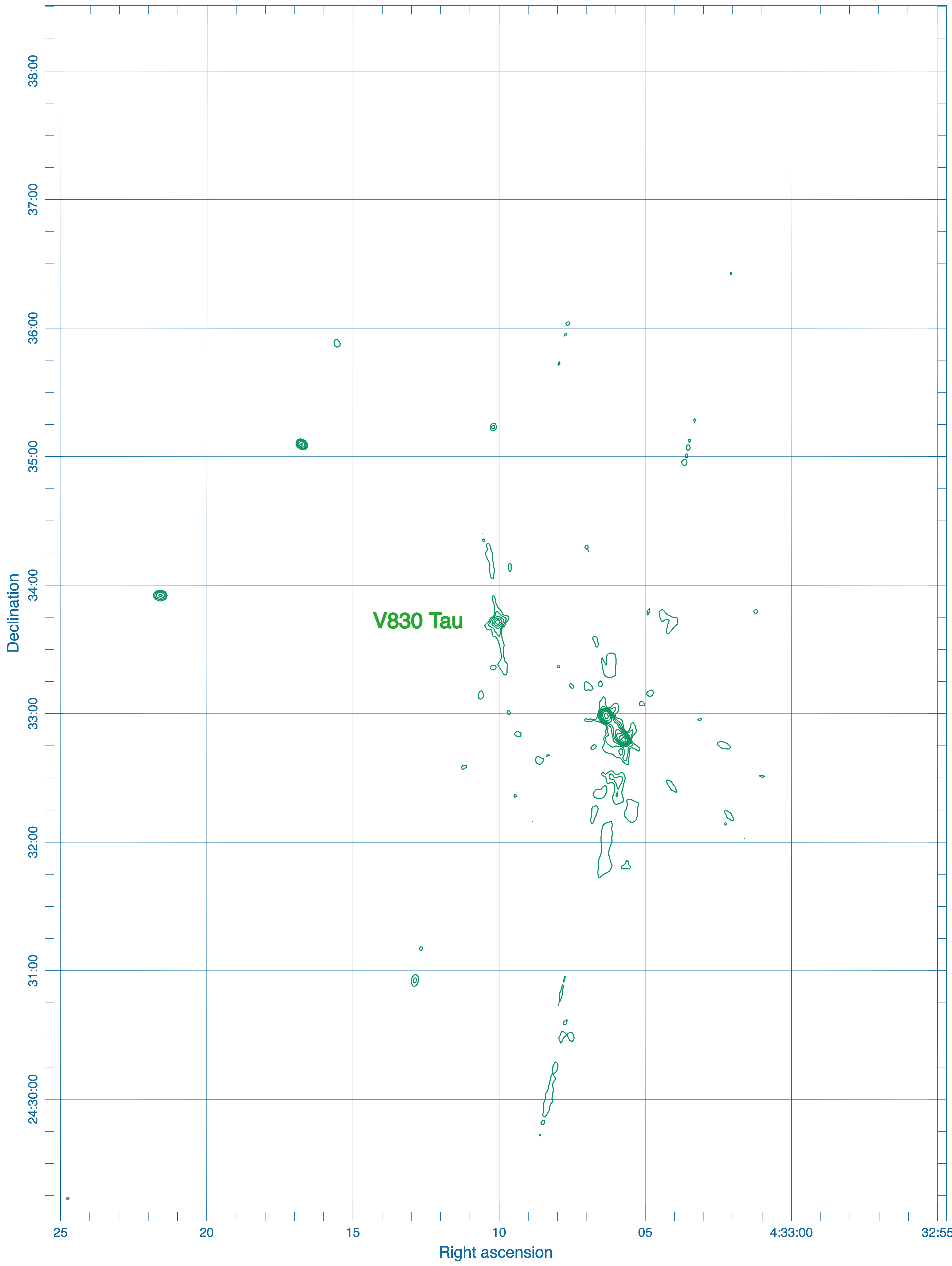}
\caption{Image of the field around V830~Tau. The cell size is 0.5''. 
 The contours are 9, 18, 36, 72, 144, 288, 576, 1150, 2300, and 4610 $\mu$Jy, with a blank-field rms of  1.8 $\mu Jy$.  
\label{fig:v830tauimage}}
\end{figure}

The observations took place as part of JVLA program 20B-268 with the array in BnA configuration. The correlator setup utilizes the C band receiver with 32 spectral windows and 16 MHz channels spanning the frequency range 3999.5-8031.5 GHz. 
Data calibration and imaging took place using CASA (version 6.5.5.21). Data from each observing block were
separately reduced, with gain, bandpass and phase calibrations applied. 
3C147 was the gain and bandpass calibrator, J0403+2600 was the phase calibrator.
Once satisfactory calibration solutions were obtained and applied to all sources, the visibilities of the target were split off, and subsequently these files were concatenated into one visibility dataset. 
The total time on source is 11.4 hours. 

Initial imaging of the field with all the data (Fig.~\ref{fig:v830tauimage}) revealed numerous sources, including several fairly bright radio galaxies.  
The rms in a blank-field region is 1.8 $\mu$Jy.
We modelled and subtracted off the visibilities from these sources. 
The target, V830 Tau, is clearly visible with a peak flux density of 590.8$\pm$4.0$\mu$Jy/beam
in total intensity and 17.1$\pm$1.2 $\mu$Jy/beam peak flux density in circularly polarized emission.

\begin{figure}
    \includegraphics[scale=0.5,bb=100 100 750 550,clip]{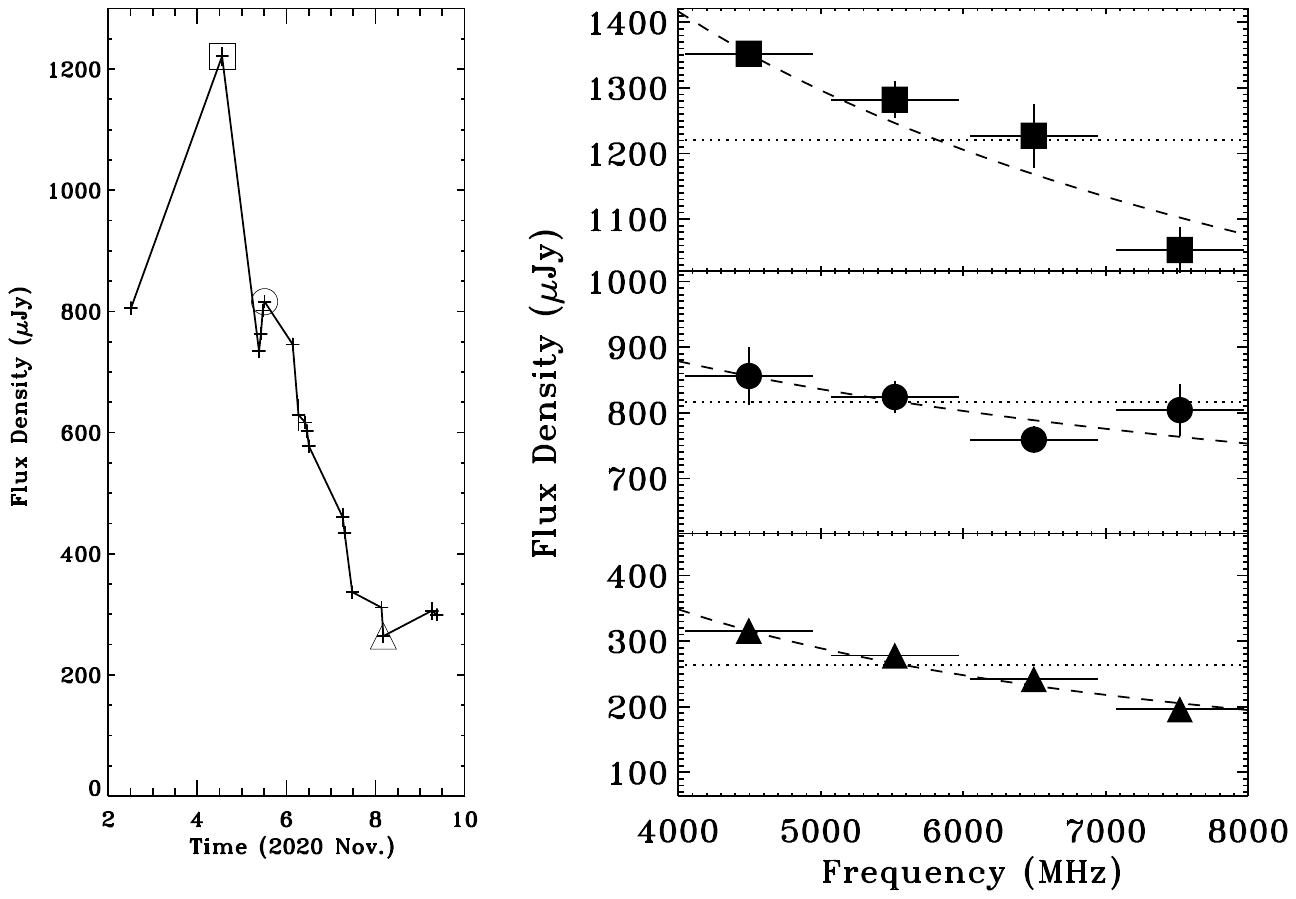}
    \caption{
    (\textit{left panel}) Light curve of V830 Tau over the observing program. Open symbols refer to the time intervals for which spectral energy distributions are shown in the right panels.
    (\textit{right panels})
    Each sub-panel shows the spectral energy distributions over the observing bandwidth, broken into 4 sub-bands, for three observing blocks whose open plotting symbol is shown in the light curve at left. 
    In each sub-panel, the dotted line shows the average flux density value over the entire bandwidth for that observation block as recorded in Table~\ref{tbl:flux}, and the dashed line shows the best-fit spectral slope. The spectral index values are tabulated in Table~\ref{tbl:flux}.
    \label{fig:lc_sed}
}
   
\end{figure}

We then proceeded to form separate images using the data from each observation block, and fitted a Gaussian at the expected position of the target determined from summing over all the observations. We included an offset in the image fitting to account for residual visibilities from the nearby radio galaxy, and fitted a Gaussian to the source.  For most of the observations, the data were consistent with a point source or a source only marginally resolved. Based on the expectation that V830 Tau is a point source at this frequency and resolution, we record the peak flux density from the image fitting for further analysis. This was done both for total intensity as well as circularly polarized emission (Stokes I and V, respectively). We formed the fractional circular polarization $\pi_{c}=V/I$ percentage by taking the ratio of the flux density in Stokes V to the flux density in Stokes I. 
These values are listed in Table~\ref{tbl:flux}.
The left panel of Figure~\ref{fig:lc_sed} displays the variation of flux 
density over the eight days of our observations. 
There is a large amount of variability, peaking at $\approx$1200 $\mu$Jy
on Nov. 4, and appearing to approach a floor in flux density of about
264 $\mu$Jy by Nov. 8.

We next repeated this exercise for four different sub-bands, to record the spectral energy distribution over the $\sim$4 GHz bandpass.  The right panels of Fig.~\ref{fig:lc_sed} show the
spectral energy distribution at the time of maximum as well as the minimum flux density.  Spectral indices were calculated,
under the assumption that $F_{\nu}\propto \nu^{\alpha}$, with 
$F_{\nu}$ the peak flux density in each sub-band, $\nu$ the observing frequency and $\alpha$ the spectral index. Spectral indices for each observation block are listed in Table~\ref{tbl:flux}.
While there is a large-scale secular change in the flux density over the course of the observation, the spectral energy distributions for each interval are generally consistent with a slightly falling trend and the circular polarization values are either undetected or indicate small amounts of polarization.

Given the strength of the source and evidence for secular variability 
over the duration of the program observations, we searched for smaller timescale variability within each observing block.  
Each target scan lasts nominally 6.9 minutes\footnote{The last scan of each observing block was shorter, at 3.4 minutes.}, and we created images and performed image analysis for each scan as described above for observation blocks. 
We calculated the ratio of each scan-specific
peak flux density to the value averaged over the observation block, and determined that the distribution of these ratios was a Gaussian with a standard deviation of 0.05.  
This is much smaller than the factor of 4.6 maximum-to-minimum flux density ratio over the days-long timescales of the program, so we instead concentrate on the values averaged over each observation block in further analysis.

\begin{table}[]
    \centering
    \begin{tabular}{llllll}
     Obs'n Block &  Start time  & Stop time & Peak Flux Density & Spectral Index &$\pi_{c}$ \\
         &    &   & ($\mu$Jy) & & (\%) \\
       \hline
  1&   11/2/2020T11:53:40.    &     11/2/2020T12:37:51.3    &  805.5$\pm$9.3    &  $-0.23\pm0.08$    & 6.8$\pm$0.4 \\  
2&  11/4/2020T12:59:35.0   &    11/4/2020T13:43:46.2     &    1221$\pm$16       &     $-0.39\pm0.05$   & 4.7$\pm$0.2 \\
3&   11/5/2020T08:54:50.  &     11/5/2020T09:38:56.3     &     734$\pm$9.7      &     $-0.24\pm0.05$      &2.2$\pm$0.5                        \\    
4& 11/5/2020T09:54:40.     &    11/5/2020T10:38:45.0     &    763$\pm$11       &      $-0.28\pm0.06$        & $<$6                \\   
 5&   11/5/2020T10:54:30.  &       11/5/2020T11:38:41.3   &   801.5$\pm$8.6        &  $-0.26\pm0.03$          &$<$1.5                    \\ 
 6& 11/5/2020T118:54:20.    &      11/5/2020T12:38:31.3    &   816$\pm$9.1        &   $-0.22\pm0.11$         &$<$1.5                 \\     
7&  11/6/2020T03:01:50.   &      11/6/2020T03:45:55.0     &    745.5$\pm$3.2       &  $-0.53\pm0.02$       &  $<$1.5               \\    
  8& 11/6/2020T06:09:20.0    &     11/6/2020T06:53:31.2      &     629$\pm$26      & $-0.32\pm0.11$        &  1.1$\pm$0.2                    \\    
9&   11/6/2020T09:47:50.0    &    11/6/2020T10:31:56.2       &       616.7$\pm$8.3    &   $-0.33\pm0.08$   &$<$1.2               \\     
10& 11/6/2020T10:47:40.    &    11/6/2020T11:31:46.2     &       602.6$\pm$5.7    &   $-0.29\pm0.04$ &  1.6$\pm$0.5 \\  
 11& 11/6/2020T11:47:30.0     &     11/6/2020T12:31:41.2     &      577.4$\pm$7.2     &  $-0.33\pm0.13$&  $<$2.1 \\
  12& 11/7/2020T06:01:15.0  &    11/7/2020T06:45:20.0      &      460$\pm$16     &   $-0.17\pm0.09$          &10.5$\pm$0.8             \\      
 13&  11/7/2020T07:01:05.0   &      11/7/2020T07:45:16.3   &   434.5$\pm$8.9        &                  $-0.46\pm0.07$       & 8.5$\pm$0.7       \\ 
14&  11/7/2020T11:07:35.0    &    11/7/2020T11:51:46.3      &      336.6$\pm$5.1     &                  $-0.44\pm0.10$    &7.5$\pm$1.1             \\  
 15& 11/8/2020T02:54:00.0    &    11/8/2020T03:38:06.2     &     311.1$\pm$8.3      &                    $-0.96\pm0.08$    &     9.6$\pm$1.2     \\   
 16&  11/8/2020T03:53:50.0  &     11/8/2020T04:37:56.2    &  264$\pm$10         &                    $-0.83\pm0.14$       &      8.1$\pm$1.5  \\  
 17&   11/9/2020T06:08:30.0  &        11/9/2020T06:52:35.0   &      306$\pm$15     &               $-0.47\pm0.15$     &  3.0$\pm$0.9          \\     
 18&  11/9/2020T08:41:05.0    &    11/9/2020T09:25:16.2       &   299$\pm$1.6        &            $-0.13\pm0.01$    &  4.2$\pm$0.5               \\
         \hline
    \end{tabular}
    \caption{Summary of flux density values and spectral indices obtained over the course of the program, for each observation block.}
    \label{tbl:flux}
\end{table}

\section{Analogous Long-Duration Flares in Active Binary Systems \label{sec:rscvn}}
Chromospherically active binary systems have long been known to exhibit large amplitude radio flares lasting days. These are systems in which the two stars are tidally locked, and exhibit much faster rotation rates than if they were single. Algol was perhaps the first such large radio flare detected, in excess of 1 Jansky \citep{Gibson1975}.
\cite{BeasleyBastian1998} reported on a large, long-duration flare event from UX~Ari over the course of nine consecutive days, with VLBA source structures consistent with interbinary emission\footnote{Notably, the radio coverage of this event, from 16-25 Nov. 1995, coincides with the decay of a luminous, long duration high energy flare seen with the Extreme Ultraviolet Explorer (EUVE) from 19-25 Nov. \citep{OstenBrown1999}.}.
\citet{osten2004} reported on flare events from the active binary HR~1099, including one lasting $\sim$two days, with higher angular resolution observations of the system discussed in \citet{Ransom2002}
and \citet{Golay2024} noting compact structures which rule out emission from an interbinary region. 
The distribution of flare durations, from hours to days, almost certainly maps to a similar distribution of flaring sizes. 
The source of these extreme flares has long been a topic of debate, whether they are localized to regions near one of the stellar surfaces, or whether the intrabinary region plays a role in connecting magnetic structures on the two stars, as in the scenario presented by 
\cite{UchidaSakurai1985}.

These systems also produce frequent and extreme high energy flares, with similar evidence for large structures. 
\citet{PandeySingh2012} analyzed  X-ray flares from several RS CVn systems, finding coronal loop lengths less than or equal to the stellar radius.
\citeauthor{PandeySingh2011} compared loop lengths derived from application of hydrodynamic models of flare decay with those derived from flares on main sequence stars using the same technique and commented that the flaring structures are larger in RS CVn binaries.
They found the loop lengths comparable to the flaring structures from pre-main-sequence stars \citep[c.f.][]{Favata2005,Getman2008} and attribute the length as an outcome of the enhanced coronal magnetic activity of the RS CVn systems.

\citet{Richards2003} presented several years' worth of continuous near-daily monitoring of three active binary systems taken with the Green Bank Interferometer at 8.3 GHz and 2.3 GHz. 
The main result presented was evidence for periodicity in the radio flaring (Algol: 48.9$\pm$1.7 days, HR1099: 120.7$\pm$3.4 days, UX Ari: 52.6$\pm$0.7 days) which was interpreted as suggestive of activity cycles. 
These data are available from https://www.gb.nrao.edu/~fghigo/ftppub/gbidata/gdata/gindex.html, and we use them in the present study for additional perspective on long-duration flares.
Details of the flux calibration, random noise, and systematic effects are discussed in the README file at the above website, as well as in \citet{Richards2003}.
In addition to the binary systems described above, there are data on TZ~CrB, a short-period (1.14d orbital period) binary system composed of two main-sequence G-type stars. While this system was only monitored for 838 days compared with 1126 for UX Ari and 1413 for HR 1099 (last column in Table~\ref{tbl:rscvn}), it provides an interesting comparison point in the discussion about bright and long duration radio flares.

\begin{table}
   \caption{Properties of three active binary systems \label{tbl:rscvn}}
    \begin{tabular}{lllllll}
        Binary & sp. type & P$_{\rm orb}$ & R$_{\star,c}^{1}$ & Sep. & dist. & $\Delta$t$_{\rm GBI}^{*}$\\
                &  &(d)           & (R$_{\odot}$) &  (R$_{\odot}$) & (pc) & (d)\\
        \hline
        UX~Ari$^{\dagger}$  &K0IV+G5V & 6.44  & 5.78 & 32.0 & 50.6 &1126\\
        HR~1099$^{\dagger}$ & K1IV+G2V&2.84 & 3.8 & 	11.5&  29.4 &1413\\
        TZ CrB$^{\ddagger}$ &F9V+G0V &1.14 & 1.21 & 5.99 & 21.1 & 838\\
        \hline
        \multicolumn{7}{l}{$^{1}$ radius of the cooler component of the binary}\\
        \multicolumn{7}{l}{$^{\dagger}$ Properties taken from \cite{Eker2008}}\\
       \multicolumn{7}{l}{$^{\ddagger}$ Properties taken from \cite{Raghavan2009}} \\
       \multicolumn{7}{l}{$^{*}$ Duration of monitoring time with Green Bank Interferometer; see text for details.} \\

    \end{tabular}
\end{table}

\begin{figure}
    \includegraphics[scale=0.5]{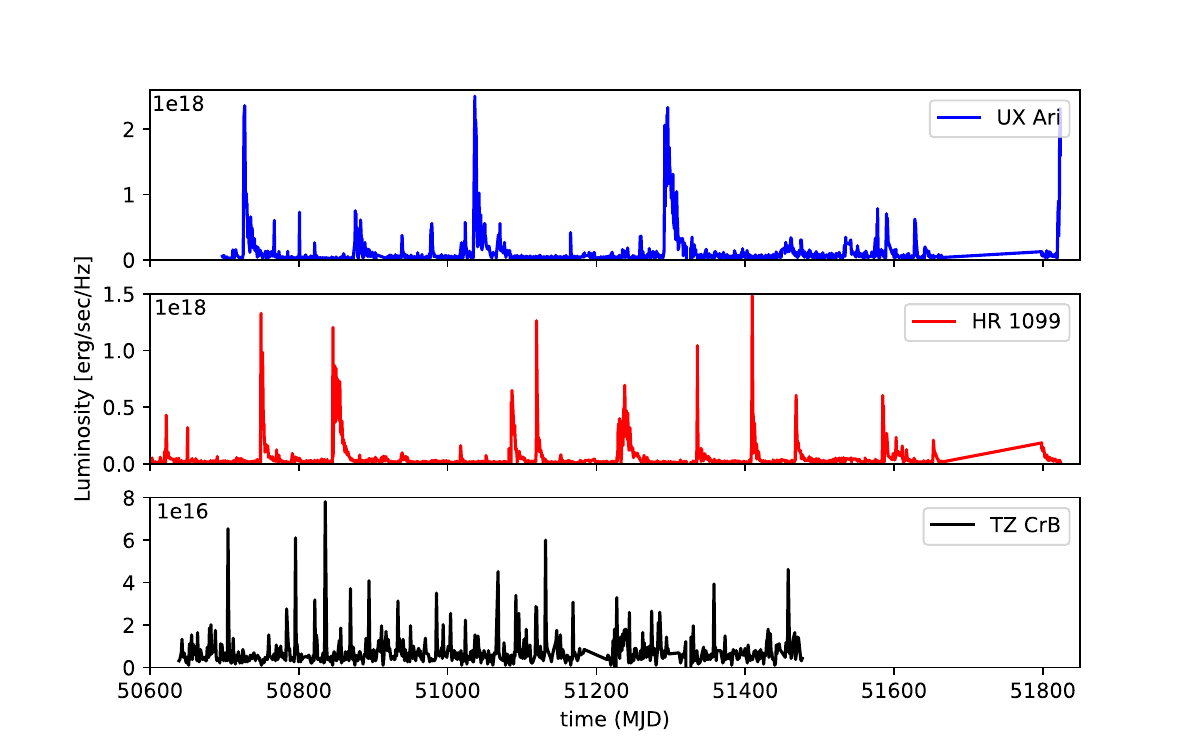}
    \includegraphics[scale=0.5]{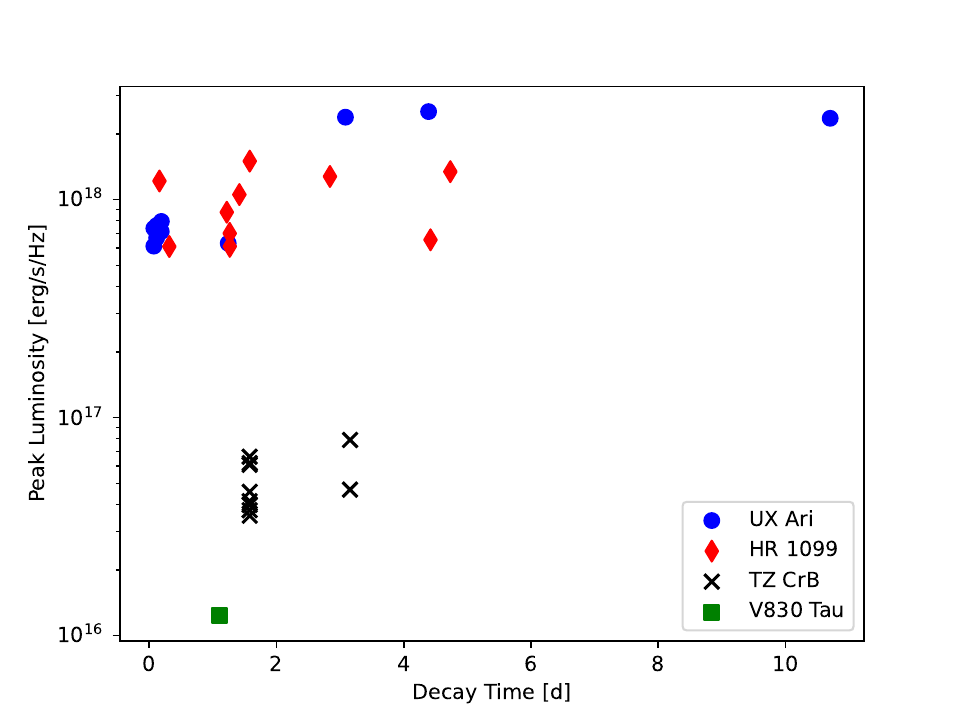}
    \caption{Left: Figure showing radio (8.3 GHz) luminosity light curves from the Green Bank Interferometer of the active binary systems UX Ari, HR 1099, and TZ CrB. Note the exponential scaling above each plot. Right: The peak radio luminosities as a function of decay time for the 10 brightest flares from each system.  UX Ari is shown as blue circles, HR 1099 as red diamonds, and TZ CrB as black X's. The green square indicates the location of the V830 Tau flare peak for comparison; see \S~\ref{sec:loops} for additional discussion.}
    \label{fig:gbi}
\end{figure}

For our analysis, we focus on UX~Ari, HR~1099, and TZ~CrB, as Algol contains evidence of mass transfer \citep{budding1989}. The binary properties are summarized in Table~\ref{tbl:rscvn} and the 8.3 GHz light curves
are shown in Figure~\ref{fig:gbi}.
Each of the stars has a distinct structure to its lightcurve.  UX Ari in particular experienced strong impulsive flares exceeding 2 $\times 10^{18}$ erg sec$^{-1}$ Hz$^{-1}$ at 8.3 GHz approximately every 300 days.  
There are usually three to ten smaller flares 
between each large flare.  
This was described as periodicities at $\sim$141.4 days and $\sim$52.6 days by \citet{Richards2003}.  
These were also the most powerful flares in terms of luminosity.  
Several flares on HR~1099 exceeded  $10^{18}$ erg sec$^{-1}$ Hz$^{-1}$. 
There was also a periodic nature to the large flares with periodicities similar to those of UX Ari, but in contrast
 there were only a few small flares in between the big events.  In contrast, flares on TZ CrB have very different characteristics, with lower peak fluxes (largest at about 8 $\times 10^{16}$ erg sec$^{-1}$ Hz$^{-1}$, 20-30 times lower than the brightest flare peaks in HR~1099 and UX~Ari), 
a noticeable flare approximately every 20 days, 
and a significant lack of long duration flares. 

We calculated the decay times of the 10 brightest flares for each of the three RS~CVn stars.
We detected flares by inspecting luminosities in the 8.3 GHz band, because the flares were strongest in this band. 
For each star, we set a detection threshold such that we detected 10 flares for each star.  
We used a peak-finding algorithm to locate peaks above this minimum threshold.
This threshold was 6.0$\times 10^{17}$ erg sec$^{-1}$ Hz$^{-1}$  for UX Ari, 5.2$\times 10^{17}$ erg sec$^{-1}$ Hz$^{-1}$ for HR 1099, and  3.5$\times 10^{16}$ erg sec$^{-1}$ Hz$^{-1}$ for TZ CrB.  
 Once this  threshold was crossed, we
considered that the flare had not ended until the
 luminosity was below 20\% of the  threshold. 
This value was chosen to try to get a complete duration of the flare while not getting confused by low-level variability/noise as the signal fell near the quiescent level. 
An approximate $e$-folding timescale was determined from the slope evaluated between the peak flare luminosity and $1/e$ of that luminosity, and the difference in time $\Delta$ t between the time of peak luminosity and the time at which the luminosity passed below $1/e$ of the peak.  For flares with a rapid decay such that the next data point in the light curve was already below 
$1/e$ of the peak, we took $\Delta t$ to be the difference in time between the two consecutive points.  
The decay time $\tau_{d}$ was then  estimated as $\tau_{d}=-e/(1-e) \Delta t$.
We note that the last flare observed on UX Ari (near MJD 51840) was not observed in its entirety and therefore excluded from the following discussion. 

The right panel of Figure~\ref{fig:gbi} displays the relationship between the peak radio luminosity and decay time, comprising ten flares from each of the three systems. 
The least luminous flares all have short decay times. This is not surprising; these are all active stars, and many small localized flares are expected. Then there is an intermediate group of flares with peak luminosities between 5$\times 10^{17}$ erg sec$^{-1}$ Hz$^{-1}$ and 3$\times 10^{18}$ erg sec$^{-1}$ Hz$^{-1}$ that can either have short or long decay times.
UX Ari has the largest separation of the three systems (as well as the largest stellar radius of the cooler binary component) and also hosts the most luminous and longest-lived flares. TZ CrB has the smallest separation of the three systems ( as well as the smallest stellar radius of the cooler binary component) and does not host long-lived flares.

\textbf{ 
} 

\section{Long duration flares as signatures of extended magnetic loops \label{sec:loops}}

\begin{figure}[h]
    \includegraphics[scale=0.4]{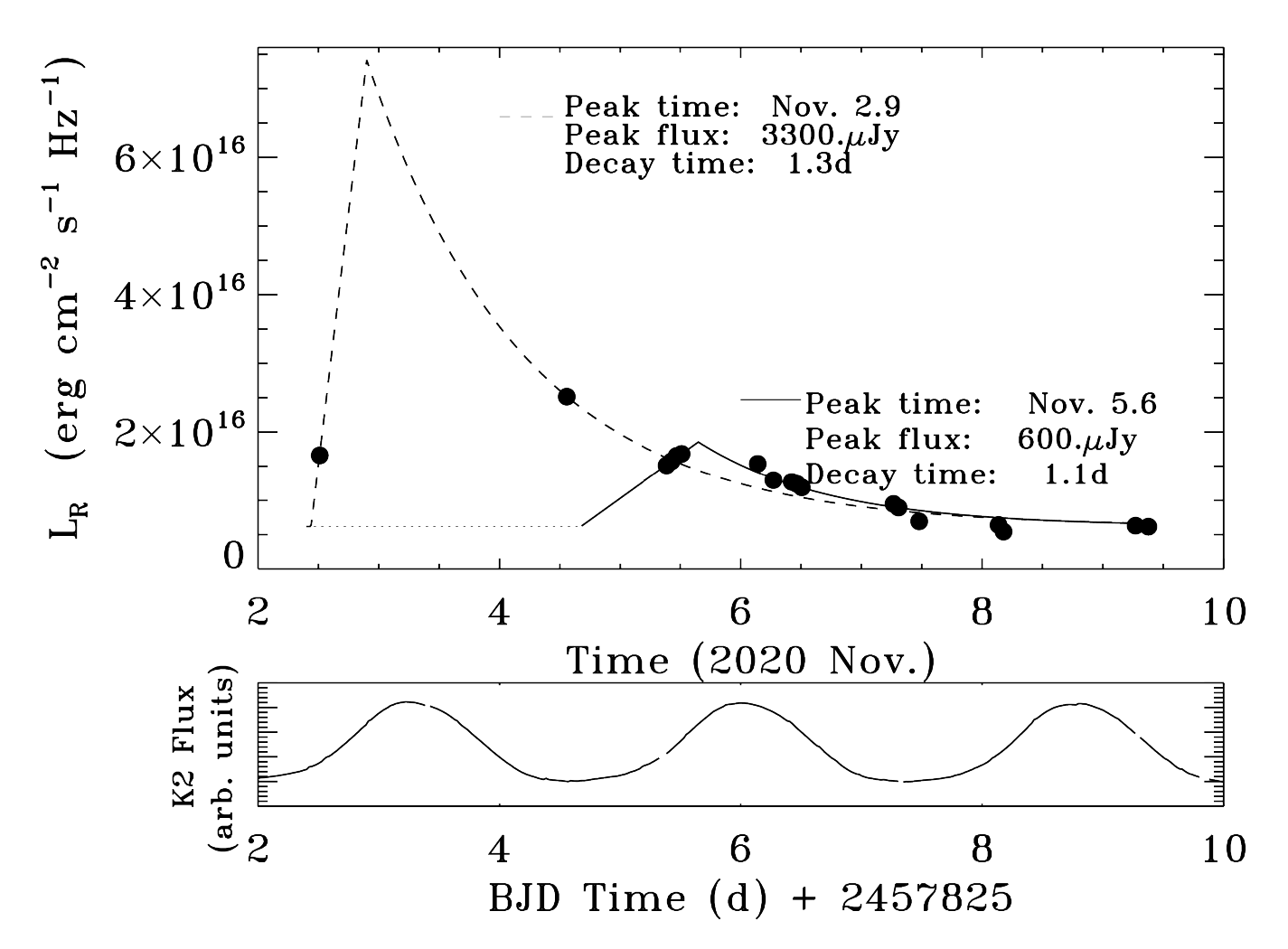}
    \caption{\textit{(top panel)} Light curve of radio luminosity in V830~Tau averaged over each observation block; see text for details. Two possible interpretations of the long-timescale secular variability are shown, with a linear rise and exponential decay in addition to a constant level (shown with a dotted line). The first scenario depicted by the dashed line suggests a peak enhancement of more than a factor of ten, while the second scenario depicted by the solid line shows a flare lasting multiple days as constrained by the data with higher observing cadence.
    \textit{(bottom panel)} Non-simultaneous optical light curve
    from K2 showing the stable 2.75 day rotation period. 
    Note that data have not been phase-shifted.
    \label{fig:flare}}
\end{figure}

We appear to have observed a luminous, long-duration flare event
on the V830 Tau system. The sparse coverage around the time
of maximum flux density prevents robust connection of this behavior 
to what is observed from  Nov. 5-9, and so there could be multiple
flaring events during this time. 
We present two scenarios in Fig.~\ref{fig:flare} -- both assume that the flaring event is characterized by a linear rise and an exponential decay. 
In one case, we constrain all of the observed data points as being part of a single long-duration luminous event. In the second scenario, we exclude the first and second data points from the event.  This implies a slightly less luminous and slightly shorter duration event.  There is also an argument for an intermediate case for a flare which peaks on or near Nov 4.5. As described above, the most conservative interpretation of the data is that the events on Nov 5 and afterwards indicate a long-lived flare
with a peak total flux density (flare and quiescence) of $\approx$ 900 $\mu$Jy.

The first two high flux density points could represent other flare enhancements, but without accompanying data we cannot constrain this further.
We note that the highest observed luminosity from V830~Tau is on par with the lowest luminosity events from TZ~CrB studied above.

For particles with an almost instantaneous injection into a closed magnetic loop, those which have a loss cone pitch angle larger than a critical value will be
mirrored in a magnetic trap, while those less than this critical value will be lost from the trap. 
The slow decay implies long-term trapping and a consequent large mirror ratio \citep{Lee2002}.
We follow the reasoning in \cite{Lee2002} identifying the decay time of the radio flare as indicative of the precipitation rate in a trap plus precipitation model, where at late times the
radio flux density proceeds as $F(t)\propto e^{-\nu t}$, with $\nu$ the precipitation rate. 
The critical pitch angle $\theta_{0}$ can be expressed as the ratio between the magnetic field strengths at the foot points ($B_{0}$) and loop top ($B_{1}$)
of a magnetic structure, according to \\
\begin{equation}
  \sin \theta_{0}=\left( \frac{B_{1}}{B_{0}} \right)^{1/2} \;\;\;
\end{equation}

which in the small angle limit reduces to $\theta_{0}=\left( \frac{B_{1}}{B_{0}} \right)^{1/2}$.
The precipitation rate $\nu$ can be written in the strong diffusion limit
as \\
\begin{equation}
    \nu = \frac{1}{2} \theta_{0}^{2} \frac{c}{H_{1}-H_{0}}
\end{equation}
where the structure extends from $H_{0}$ (with associated field strength $B_{0}$) to $H_{1}$ (with associated field strength $B_{1}$)
above the surface of the star. 
\citet{massi2006} used this geometry to argue for the existence of helmet streamers in the coronae of V773 Tau, which displays periodically recurring short-timescale mm flares during times of periastron passage of the two stars. 
For a dipole configuration, $B_{1}/B_{0}$=$\left( \frac{H_{0}}{H_{1}} \right)^{3}$.
In the \citeauthor{massi2006} scenario, the value of $H_{0}$ need not be set at the stellar radius \citep{Endeve2004}. 
To match the observed exponential decay of the flux density light curve, the precipitation rate will be equal to the inverse of the decay timescale, giving 
$\nu = 1/\tau{_d}$. 
This results in a relationship between the decay timescale of the flare and the two distances, \\
\begin{equation}
    \frac{1}{\tau_{d}}= \frac{1}{2} \left( \frac{H_{0}}{H_{1}} \right)^{3} \frac{c}{H_{1}-H_{0}} \;\;\; .
\end{equation}
This can be rearranged to give a function f($H_{0},H_{1}$)\\
\begin{equation}
   f(H_{0},H_{1}) = \frac{2 (H_{1}-H_{0})}{c \tau_{d}} -\left( \frac{H_{0}}{H_{1}} \right)^{3}
   \label{eqn:h0h1}
\end{equation}
and zeroes of the function, and thus appropriate values of $H_{1}$, can be determined once
$H_{0}$ is specified.

\begin{figure}[h]
\includegraphics[scale=0.5]{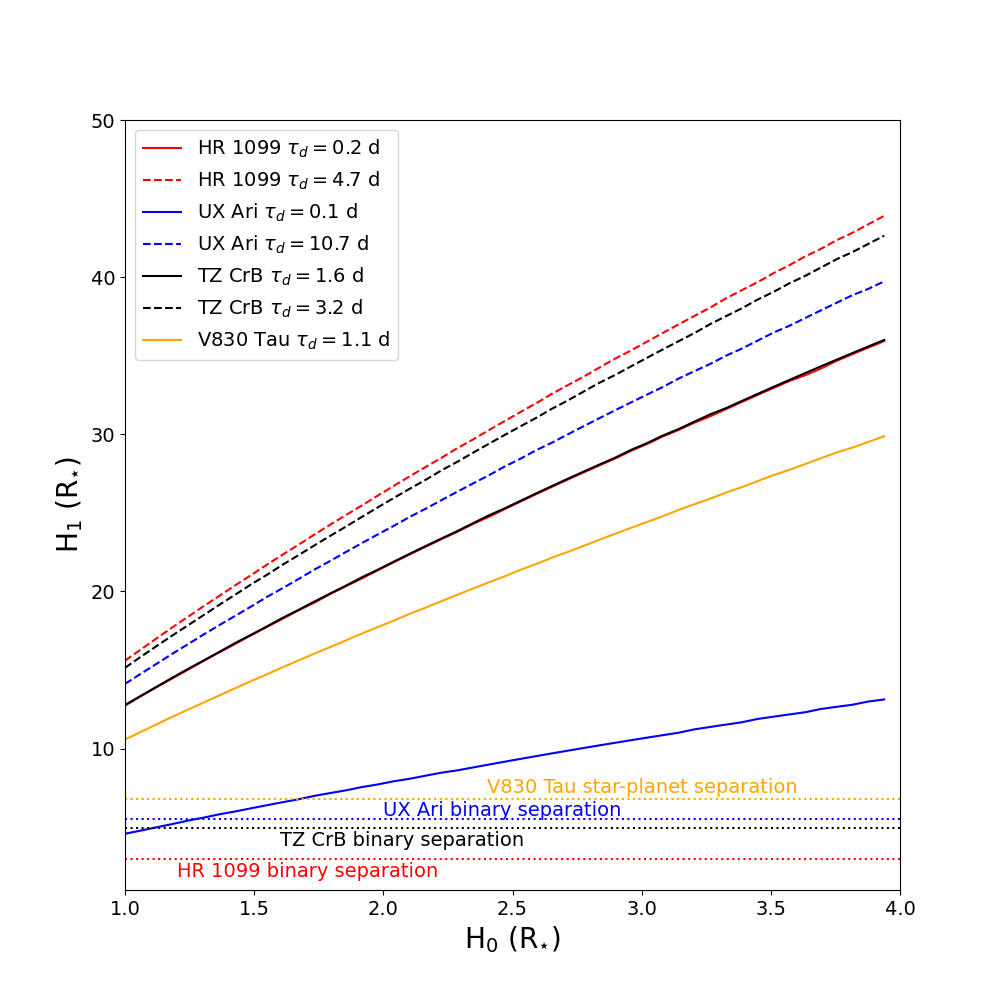}
\caption{ Constraint on loop sizes using the helmet streamer methodology of \citet{massi2006}. Each point in a curve is a pair of lower and upper heights (H$_{0}$,H$_{1}$) which satisfies equation \ref{eqn:h0h1} for the given flare timescale and stellar radius; units are the stellar radius of the cool component of the stellar binary or the radius of V830 Tau. 
The shortest and longest decay timescales for the three active binaries UX~Ari, HR~1099, and TZ~CrB are described in \S ~\ref{sec:rscvn}, shown respectively in solid and dashed lines, while flare decay time for V830 Tau is taken from discussion in Section~\ref{sec:loops}. 
Horizontal dotted lines indicate the binary separation in units of the stellar radius.
\label{fig:loops}}
\end{figure}

With this framework in mind, we turn to the V830 flare, and use scenario 2 above, which constrains the decay timescale to be $\tau_{d}=1.1 $d (Fig.~\ref{fig:flare})
and the flare duration to be roughly 4.5 days.  
To this we add the constraints on flare decays from the active binary systems with long-term monitoring described in \S~\ref{sec:rscvn}. 
Fig.~\ref{fig:loops} displays the results, using the minimum and maximum flare decays for the active binary systems, and the V830 Tau flare decay.
For these purposes the reference value H$_{0}$=1 is assumed to be at stellar surface.

The longest flares can have lengths extending up to 15 $R_{\star}$  if they originate at the stellar surface, and even larger if the lower height is raised above the stellar surface, as in the case of a helmet streamer. 
Conversely, even the shortest flares would have heights extending to $\approx$4.5 R$_{\star}$ if they originate at the stellar surface.
For all but the shortest decay time flare seen on UX~Ari, the implied loop height assuming $H_{0}=1$ exceeds the binary separation for UX Ari, HR 1099, and TZ CrB, and the star-planet separation for V830 Tau.

\section{Discussion}
The two main discussion items are the nature of the long duration flare, and whether it is possible to make any connection to magnetospheric interactions between the star and its disputed planet.    These are discussed individually in the following sub-sections. 

A major assumption being made in this analysis is that the structure producing the radio brightening occurs in one monolithic magnetic loop. 
There have been many studies inferring large flaring loops \citep[such as][who analyzed a large flare on the RS CVn GT Mus and found a loop length of 60 $R_\odot$]{Sasaki2021}.
Only long baseline radio interferometry has the ability to spatially resolve the emission, and results have revealed the radius of magnetospheres around young stars to be of order 6R$_{\star}$ \citep{Torres2012}. 
While resolved imaging of the solar corona does reveal loop-shaped structures involved in flaring \citep{reale2014}, another scenario is also observed, in which an arcade of closely-connected loops are sequentially triggered to produce the overall flare event \citep[such as the famous Bastille Day event, described in ][]{bastilleday2001}.
Either of these scenarios (a single loop extending many times the length of the star, or sequential triggering of an arcade for longer that the stellar rotation period) challenges the understanding of coronal structures from the solar context; see also the discussion in chapter 12.1 in  \citet{kowalski2024}.

The circular polarization properties of V830 Tau are similar to those reported for a few other T Tauri stars.
 \citet{white1992}  and \citet{phillips1996} noted a circularly polarized fraction of 2-4\% of total intensity at cm wavelengths, while the range
 seen here is $<$10\%, see Table~\ref{tbl:flux}.
Given the evidence presented here for large structures,
this would be a good system to probe for evidence of linear polarization in analogy with V773 Tau. 

We note in passing that \citet{Gao2020} have described potential signatures of exoplanetary synchrotron radio bursts, detailing the light curve and frequency shift as key features of planetary synchrotron radio bursts. 
The timescales involved pertain to the time for relativistic electrons to travel from the star to the planet ($\approx$ 30-36 s for 300 keV - 1 MeV particles for the case of V830~Tau) and do not match with the timescales observed here.
Likewise, we see no evidence for frequency shifts over the course of our 4 GHz receiver band, although from inspection of their Figure~3, the timescales over which such a shift might take place would be less than a second at GHz frequencies. 

\subsection{On the Nature of the Long-Duration Flares}
The analysis in \S 4 from the JVLA light curve of V830 Tau, as well as extensive radio monitoring of a handful of active binary systems, shows that long-duration flares can plausibly be explained as originating in large structures.  
 However, there is an inherent degeneracy between the size of the cooler component of the binary system and the intra-binary separation in the types of objects that are used for comparison, as seen from examination of Table~\ref{tbl:rscvn}. 
In this admittedly small sample, the binary system with the largest separation between the two stars, UX~Ari, at 32 $R_{\odot}$, also has the largest stellar radius of the cooler component, at 5.78 $R_{\odot}$, while the binary system with the smallest separation, TZ~CrB, at 5.99 $R_{\odot}$, has the smallest stellar radius of the cooler component, at 1.21 $R_{\odot}$.

There are categories of single active evolved stars that display flaring activity, some of which are also of extended duration. 
\citet{Slee1987} reported on the ultrafast rotating single giant star
YY~Men \citep[K1~III, R$\sim$12R$_{\odot}$;][]{Strassmeier2009}  with a days-long radio outburst.
\citet{Ayres1999} observed a high energy flare from the yellow giant $\mu$~Vel \citep[G6~III, R$\sim$13R$_{\odot}$;][]{Mullan2006} which lasted roughly two days, and
\citet{Ayres2001} demonstrated the ubiquity of days-long extreme ultraviolet flaring activity on the single clump giant $\beta$ Cet \citep[K0~III, R$\sim$17.5 R$_{\odot}$][]{Reffert2015}. The discussion in \citet{Ayres1999} uses the long duration as a signpost of the flare occurring in a ``significantly elongated structure."
These objects are all different:
$\mu$Vel is an intermediate mass star whose path away from the main sequence has placed it near the Hertzsprung gap in the "rapid breaking zone"; $\beta$ Ceti is an intermediate mass ``clump'' giant which has already ascended the giant branch and undergone a helium flash. 
YY~Men is considered to be an FK-Com-type system --
apparently single, but with its rapid rotation occurring as a result of the engulfment of a former companion. 
Rapid rotation appears to be the key to the extremes of magnetic activity presented here, and the long durations are indicative of large coronal structures.
There appears to be a paucity of systematic studies, particularly of radio flaring in post-main sequence contact binary systems, which would break this degeneracy. 
We cannot thus causally connect the long-duration flare to the existence of a companion.

V830~Tau is a pre-main sequence star and has a stellar radius larger than solar (1.8 R$_{\odot}$), while its companion, the Jupiter-mass exoplanet V830 Tau b, is
disputed \citep{Damasso2020}.  The degeneracy noted above between binary separation and stellar radius, therefore, means that we cannot argue for the existence of V830 Tau b based on the observation of the days-long radio flare.
We note that based on Figure~\ref{fig:loops}, even a helmet streamer
originating on the surface of V830 Tau would extend
out to $\sim$10 R$_{\star}$, which is past the location of the putative planet, and could be used to argue for magnetospheric interaction.

\subsection{On the Expected Star-Planet Interaction Power}
Recurring flares at or near the orbital period of the system would bolster claims for the existence of the planet and present an analogy to the interacting magnetospheres scenario presented in \citet{massi2002}.  
If the emission originates close to the stellar surface, the modulation would be at the orbital period, whereas emission from an extended stable loop-like feature would recur with a beat period between the rotational and orbital periods.
It can be shown, by modeling the three-dimensional magnetic field strength distribution of V830~Tau, that a planet orbiting at $\approx$6 R$_{\star}$ would cross the Alfv\'{e}n surface, and could be expected to produce reconnection events at these times \citep{VidottoDonati2017}. 
\citet{Ilin2024} analyzed optical flare clustering in planet-hosting stars, finding that the clustering of flares which occur in phase with the planet's orbit is expected to become more pronounced as the power of the interaction increases. 
Using the formulation for star-planet magnetic interaction, in which the magnetic fields of the star and planet stretch and break \citep{Lanza2012}
following the discussion in \citeauthor{Ilin2024} (their equation 6)
for non-linear and axisymmetric magnetic fields,
we find about 5$\times$10$^{27}$ erg s$^{-1}$ of power
generated for a magnetized planet with a field strength of 1 G.
This assumes a magnetic field strength of 300 G as deduced for V830 Tau from \citet{Donati2015},
 a radius of the stellar radius, and a planetary radius of 1.8 Jupiter radii (R$_{J}$).
These assumptions are made because only axisymmetric fields are capable of producing modulation of flaring activity phased with the planetary orbit. 
We use an inflated value of planetary radius as evolutionary models for young giant planets \citep{baraffe2003} at ages of a few Myr indicate values between 1.5-2 R$_{J}$ for irradiated and non-irradiated planets.
Comparing this with \citeauthor{Ilin2024} Figure~7 and Table~3, we see that this system should exhibit a rather high interaction power, just below only HIP~67522 and TAP~26. 
This suggests that repeated radio monitoring could be fruitful in searching for additional flaring events at the same orbital phase
to bolster the argument for the existence of the planet around V830 Tau.
We note that recently \citet{Ilin2025} reported on 15 optical flares on the star HIP 67522 which cluster near the transit phase of the innermost planet, indicating persistent star-planet interaction in the system.

\section{Conclusions}
We reported on a large, long-duration flare on a 2.5 Myr star with a purported planet. 
For young stars, this long timescale monitoring of several consecutive days at GHz frequency is a new parameter space, as previously only inter-epoch variability over weeks-months, or extreme short-duration flares within a few hours had been noted \citep{Forbrich2017}. 
We examined this long-duration flare in the context of a long-term monitoring 
campaign of a few highly active binary systems and demonstrated that long-duration flares preferentially occur on evolved stars and 
stars in binary systems with orbits of 2 days or longer.
Under the interpretation that the flare timescales are controlled by 
the rate at which trapped electrons leave the magnetic trap, we estimate
loop heights, and demonstrate that longer duration flares imply longer loops than shorter duration flares. 

The extended duration over which V830 Tau was monitored ended up being ideal to detect this long-duration flare. The program was initially set up to probe the idea of periodically occurring flares at the orbital period of the system.
Given our inability to separate the effects of large stellar surface area from 
interbinary separation in controlling long-duration flares, we cannot use these data to argue for the existence of a Jupiter-mass exoplanet orbiting around V830 Tau. 
However, given the high expected power of star-planet interaction, and the disputed nature of the planet, this system is a prominent target to be monitored at radio wavelengths to investigate flare timing relative to the disputed planet's orbital period.  This could be an independent mechanism for confirming the presence of planets around magnetically active young stars. 

Based on these results, future radio projects capable of repeated monitoring of a sample of objects over the course of several years, similar to what the Green Bank Interferometer did, clearly have a huge amount of discovery space. 
There is a prospect for such night-time monitoring with a solar-dedicated
radio array \citep{Gary2023} currently being designed. 

\begin{acknowledgments}
    RAO wishes to thank informal discussions with Geoff Bower and Sergio Dzib at the ngVLA Follow the Monarchs Meeting in November 2024.
    The Green Bank Interferometer is a facility of the National Science Foundation operated by the NRAO in support of NASA High Energy Astrophysics programs. 
    S.J.W. acknowledges support from the Smithsonian Institution and the Chandra X-ray Center through NASA contract NAS8-03060.
\end{acknowledgments}

\facilities{VLA,GBI}

\software{CASA \citep{casaref2022}, CARTA \citep{cartaref}}





{}

\end{document}